\documentstyle[aps,eqsecnum,epsf]{revtex}

\begin{document}

\baselineskip=16pt

%\draft

\title{Scenario for Ultrarelativistic Nuclear Collisions:
  V.~ Onset of Deconfinement\\ (How the Nuclei Get Unbound). }
\author{ A.  Makhlin }
\address{Department of Physics and Astronomy, Wayne State University, 
Detroit, MI 48202}
\date{September 6, 2000}
%\date{\today}
\maketitle
\begin{abstract}
We consider a Euclidean extension of the wedge form of Hamiltonian
dynamics, which explicitly accounts for the strong localization of the
first interaction in nuclear collisions. A new principle of the analytic
continuation via the tetrad vector is introduced. We discover the
existence of self-dual solutions with short life-times
(ephemerons) and conjecture that these vacuum fluctuations can lower the
Euclidean action of the system of the colliding nuclei, thus enforcing a
breakdown of the nuclei coherence. We suggest that the
ephemerons can be identified with the gluons-partons, which are resolved
in high-energy nuclear collisions.
\end{abstract}

\section{Introduction}
\label{sec:S0}

In this paper, we continue our study of the scenario of ultrarelativistic
heavy ion collisions. By scenario we mean the continuous temporal
sequence of  stages, smoothly developing one into another. These stages
are different only in the respect, that each of them is characterized by
its individual {\em optimal set} of  normal modes. Posed in this way, the
problem of the scenario falls out of the jurisdiction of  scattering
theory, and makes a powerful formalism of the S-matrix inapplicable.
Instead, we suggested to use the formalism of quantum field kinetics
(QFK) \cite{QFK} and to treat the scenario as a problem of an inclusive
measurement. In Ref. \cite{QGD} and the previous papers of this cycle
\cite{tev,gqm,wdg,fse} (further quoted as papers [I - IV]), we focused on
different aspects of the scenario. We began, in Ref.~\cite{QGD}, with the
issue of the temporal evolution of quantum fluctuations consistent with a
given inclusive probe and revealed its identity with the well-known QCD
evolution. This perspective has been broadened in paper [I] and brought
us to the conclusion that the  dense system of quarks and gluons, which is
commonly associated with the quark-gluon plasma (QGP), can be formed only
{\em in a single quantum transition}. This conclusion is based on the
fact that the scale of the entire process is associated with the
properties of the final state. In papers [II] and [III], we studied
the geometrical properties of the collision process, concentrating on the
localization of the initial interaction, which is a consequence of the
{\em finite size} of the colliding nuclei. The Lorentz contraction
confines the whole process within the past and the future domains of
their intersection. The existence of the {\em macroscopic} light-cone
limits was accepted as the classical boundary condition for all quantum
fields that emerge in the collision process, leading  to a new {\em wedge
form} of Hamiltonian dynamics, which employs the proper time $\tau$ as a
Hamiltonian time of the evolution. Finally, in paper [IV], using the
framework of the wedge dynamics, we computed the effective mass of a soft
quark propagating through the expanding background of hard partons. The
leading interaction, which is responsible for the effective mass, appeared
to be the chromo-magneto-static interaction of the color currents flowing
in the rapidity direction. In paper [II], such currents were shown to be
an intrinsic property of the states of wedge dynamics.

That part of the study was motivated by the problem of entropy
production in ultrarelativistic nuclear collisions. Namely, we had to
find a basis of states which is suitable for the explicit computation of
the entropy. It turned out, that the states of this basis are formed
gradually. The method of QFK which, in its essence, is the calculus of
time-dependent Heisenberg observables, has been designed as an {\em ad
hoc} tool adequate to that part of the problem.
The QFK reduces the problem of an inclusive measurement\footnote{ As we
emphasized in paper [II], the collective interactions in the final state
can be viewed as an inclusive measurement. QGP is a collective detector
for the quark and gluon modes after the first interaction in nuclear
collision.} to the study of the possible field configurations
(fluctuations) that could develop before the measurement  and are
compatible with a given probe. Thus, this is a problem of real-time
fluctuations limited by some observable at the end of the evolution, and
driven by pQCD.  In order to comply with the
causality principle, these unobserved fluctuations must be renormalized
in such a way, that {\em before} the localized interaction of the
inclusive measurement indeed happens, the decomposition of the nuclei in
terms of the final-state modes can be only virtual.
The main object, that QFK is dealing with, is the inclusive cross section
(as compared to the exclusive amplitude of the S-matrix formalism). The
integral equations of the QCD evolution sum all the {\em exclusive
probabilities} over a complete set of the unobserved {\em states}.
Therefore, the perturbative expansion of the inclusive cross section
includes the on-mass-shell correlators corresponding to these {\em
states}. As it was shown in papers \cite{QGD,tev}, their presence allows
one to establish the temporal order of the accomplished subprocesses.

The nature of the mechanism that is responsible for the initial breakdown
of the nuclei coherence is not yet clarified. We do not know: how the QCD
interactions that make color charges visible are suddenly switched on, if
there were any time dependent fluctuations of color charges before the
nuclei has physically intersected,\footnote{In the case of two neutral
atoms bound by electromagnetic forces, the time-dependent
fluctuations of the charge density do exist and lead to the van-der-Vaals
interaction between the atoms. This kind of interaction between the
hadrons is not known.} and what is the geometry of the primordial colored
objects. The most popular parton model (in all its modifications)
implicitly relies on the impulse approximation and a time delay in the
infinite momentum frame, which does not seem to be fully consistent with
the picture of the gradually formed final states. In this paper, we put
forward a hypothesis that can lay the footing for a more detailed study of
this issue.

Our key observation is that before the moment in time when the nuclei
overlapped geometrically, nothing except the nuclei themselves can be
qualified as quantum states, and all conceivable fluctuations  should
be treated as the classical fields that coherently add up and form the
two nuclei. Hence, addressing the QCD evolution before this moment,
we must study not the inclusive probability,
$${\cal P}(A_i + B_i \to a_f + X) =\sum_{X}
 |\langle X |a_f S A_i^\dag B_i^\dag |0\rangle|^2 , $$
which sums up various contributions from the physical final states,
but the inclusive amplitude,
$${\cal S}(A_i + B_i \to a_f) =\int dxdydz \varphi_f(x)\Delta^{-1}(x)
\Phi_A(y)\Phi_B(z) \ll \int e^{iS}\varphi(x)J_A(y)J_B(z)
{\cal D}A{\cal D}\psi{\cal D}\bar{\psi} \gg ,$$
where $\varphi_f(x)$ is the wave function of the inclusive probe and the
operator $\Delta^{-1}(x)$ has appeared as the result of the LSZ
reduction. The wave functions $\Phi_{A,B}$ correspond to the initial
hadronic states and the phenomenological vertices $J_{A,B}$ connect them
with the quark and gluon fields. The functional integration can be
effective only if the newly emerging fields have a finite action and can
be treated semi-classically in Euclidean space. Then, the symbol
$\ll\cdot\cdot\cdot\gg$ stands for the average over an ensemble of these
classical fields. As a matter of fact, we are looking for a special kind
of Euclidean vacuum fluctuations which could be active in the closest
proximity of the collision point and become pure gauge fields outside an
actual ``reaction zone''. We expect that these fluctuations, being added
to the already known ones, will lower the Euclidean action and force the
conversion of the nuclei into a collective quark-gluon system.
In order to pose this problem in a consistent way, we must create a
framework, where the final states of the QCD evolution (previously
treated in the scope of QFK) will have a recognizable image among the
other Euclidean fluctuations and could be compared with instantons
\footnote{In this paper, we accept the point of view that the properties
of hadrons can be described by the ``propagation'' of quark correlators
through the instanton liquid \cite{ShuSch}, since this approach relies on
the most direct implementation of the least action principle. Perhaps,
not every reader will take this for granted. A conceivable alternative
(which is closer to the standard OPE-based approach \cite{SVZ})  would be
to study the competition between the transient field configurations and
the stationary QCD condensates.}. An example of such a {\em transient~}
fluctuation is described in Sec.~\ref{sec:S2}.  We indeed  find the
Euclidean fluctuations, which have the following properties:\\ (A) ~They
can ``compete'' with the well known fluctuations (instantons) in the
Euclidean vacuum which provide the integrity of the isolated hadrons and
nuclei;\\ (B) ~They can be resolved only due to the specific geometry of
the collision.

Acting in this way, we may hope to bridge the gap between the QFK and the
S-matrix scattering theory, and to approach the problem of the transient
process, as it was posed in the papers [I] and [II], from the Euclidean
side. These two approaches are not mutually exclusive, they just answer
different questions. Indeed, the QFK allows one to study how the final
states (which can be {\em defined} only in Minkowski space) are formed.
Then, these states can be continued to the Euclidean space as
fluctuations, and one can ask if the presence of these fluctuation can
lower the action of the whole system. To make the discussion more
consistent, we must specify
the environment where one can discuss the conditions (A) and (B).

\subsection{The Euclidean field theory.}
\label{subsec:S1a}

The most important thing one should remember when addressing Euclidean
field theory (EFT) is that the EFT is just the S-matrix scattering
theory, rewritten in a specific way. We have to explain this point in
some detail, since in most of the modern applications of EFT, it is
shaded. The design of EFT includes several steps:

1. The S-matrix amplitude is written down in Minkowski space. This is
unavoidable, since only in this way can we define the initial and final
{\em states}. (The states must be normalized on some space-like
hyper-surfaces  which exist only in the Minkowski world. Neither S-matrix,
nor the LSZ reduction formulae can be derived in Euclidean space.)

2. After an S-matrix element is transformed into the momentum
representation, the integrals over the real energies corresponding to the
internal momenta $k^\mu$ are identically transformed to the integrals
over the imaginary axis, $k^0_{M} \to i k^0_{E}$. This procedure is known
as Wick rotation. It moves the integration path away from the poles
located near the real axis. Practical convenience of this strategy of
computing the scattering amplitudes is the improved convergence of the
integrals. The Wick rotation is possible solely because the poles of the
T-ordered Feynman propagators are located in the second and fourth
quadrants of the plane of complex energy. In the course of Wick rotation,
the integration path never crosses the poles. Clearly, this step is not an
analytic continuation. If the external momenta $p^\mu$ are considered
off-mass-shell, then the external energies become the independent
arguments. The Feynman amplitude is computed on the imaginary axis of
energy $p^0$, being analytically continued to real axis at the end of
calculations.

3. The next step is to introduce an auxiliary ``coordinate
representation'' by means of the formal Fourier transform from the
Euclidean momenta to Euclidean coordinates. Acting in this way, we
translate the oscillating Minkowski propagators into their exponential
Euclidean counterparts (thus preserving the previously gained advantage
of rapid convergence). The new object still belongs to the S-matrix
theory in Minkowski space. This is the same matrix element labeled by the
quantum numbers of in- and out-particles. Finally, we can formulate a set
of mnemonic rules (resembling the rules of a real field theory) that
allow one to directly generate the results of the previous mathematical
manipulations and avoiding explicit reference to the Minkowski space.
This is what we eventually call Euclidean field theory. At the classical
level, it includes the new metric, new definitions of vectors and
spinors, new action, new equations of motion that minimize the new
action, etc.  In any of these representations, we can pass over to the
path-integral representation of the truncated amplitudes, and integrate
over the field configurations that satisfy the artificial Euclidean
equations of motion.   An
important new element that is found in the context of the Euclidean QCD
is that the auxiliary Euclidean equations of motion have topological
solutions, instantons, which have no real analog in the Minkowski world.
However, these objects make a quantitative description of the
properties of stable hadrons possible \cite{Schafer}.

A few remarks are in order:

1. Quantum mechanics strictly prohibits any observation of the
intermediate dynamics of an exclusive scattering process. None of the
intermediate Minkowski momenta or Minkowski coordinates, (as well as
Euclidean momenta or Euclidean coordinates) can be measured. There is no
``earlier'' and ``later'' in the Euclidean coordinate picture in exactly
the same way as there are no earlier and later in Minkowski description
of the interior of the S-matrix amplitude.

2. Up until now, the instanton-based calculations proved to be successful
in a relatively narrow class of problems, which  can be reduced to the
forward scattering (i.e., propagation) of a single hadron. In this case,
we can connect the physical hadronic observables and the QCD degrees of
freedom using the phenomenological quark-hadron vertices (Ioffe
currents). Though it may be very difficult to do it practically, there is
no doubt, that the method must also describe the forward scattering of a
single nucleus.

3. The Ioffe currents are color neutral and there is no colored
observables in the entire problem. This fact is remarkably reflected in
the unique polarization properties of instanton fluctuations;  Nature
``hides'' the color in a very elegant way, by identifying directions in
the color space with the directions in the non-observable Euclidean
geometric background. Once again, this is a consequence of the quantum
mechanical veto on any intrusion into the intermediate stage of an
exclusive scattering process.

4. None of the problems of real-time evolution, where causality
prescribes a certain order of interactions, e.g. the temporal dynamics of
inclusive processes, can be reduced to the EFT. Formally, the retarded
propagators just do not allow for the Wick rotation.  Physically, the
establishing of a certain temporal order requires some
measurement.\footnote{A detail discussion of the connection between an
observable and the temporal order in the inclusive process is given in
paper [I] of this cycle.} As a consequence, the inclusive measurements
(including the evolution equations that describe these measurements)
cannot be related to the instanton physics.

\subsection{Geometry of the collision and the field theory.}
\label{subsec:S1b}

The last conclusion of the previous section may sound too pessimistic,
since it seems to leave no hope for the establishing any connection
between the high-energy processes and the non-perturbative physics of the
hadronic world. We argue that this connection can be re-established due
to the finite size of the colliding nuclei, which allows one to impose
classical boundary conditions on the dynamics of {\em all} quantum
fields. These new boundary conditions create a  framework of the wedge
dynamics (proposed in papers [I] and [II]). This approach was designed in
order to solve the problem of the evolution of observables (like the
inclusive distributions of partons-plasmons) in ultrarelativistic nuclear
collisions and to give a physically motivated definition of the final states in
this process. Now we argue that the previously found final states
(corresponding to a strongly localized interaction of the two nuclei)  can
be used to pose an S-matrix scattering problem.  The latter can be
translated into the path-integral language, and one can look for the
field configurations that minimize the action of the Euclidean wedge
dynamics. To ensure the applicability of the wedge dynamics in the
theoretical analysis of the scattering problem, one has to select an
ensemble of events with the widest rapidity plateau. The properties of
the physical states in this dynamics were studied in papers [II] and
[III].

In order access the Euclidean domain of nuclear collisions, we shall keep
in mind an hypothetical S-matrix amplitude in which the connection
between the initial states and the fundamental fields of QCD is provided
by something like giant Ioffe currents of the multi-nucleon systems.
Since we aim at the scenario of a nuclear collision, we must view the
intermediate collective modes of the expanding quark-gluon matter as the
final states. Using these states, we can pose a scattering problem, which
can be then analytically continued to Euclidean variables and given a
path-integral representation. It is plausible that, in the first
approximation, the problem still can be reduced to the ``propagation'' of
the quark currents through an ensemble of the Euclidean configurations of
the gluon field. In wedge dynamics, the external variables of this
problem are the rapidities of the incoming nuclei, and the rapidities and
the transverse momenta of the final state modes. (As we shall see later,
the rule of correspondence between the Minkowski and Euclidean rapidities
reads as $~\theta_{\scriptscriptstyle M}\to i\theta_{\scriptscriptstyle
E}$.) In this geometrical context, we will be able to describe
propagation of {\em two nuclei} using {\em the same Euclidean variables}.
Furthermore,  we indeed will find a new class of self-dual solutions that
minimize the Euclidean action and are active only in the closest
proximity of the collision center. This result opens an opportunity to
approach the binding interactions in the nuclei and the high-energy
process of the breakdown of their coherence from the same point of view.

\bigskip

The rest of the paper is organized as follows. In Sec.~\ref{sec:S1}, we
review the geometric background of the wedge dynamics in Minkowski space
and pass over to its Euclidean counterpart by means of the analytic
continuation of the time-like tetrad vector. In Sec.~\ref{sec:S2}, we
explicitly find the self-dual solutions of the Euclidean wedge dynamics
that appear to be the short-lived eigenstates of the $SU(2)$ color
matrix $\sigma^3$ (ephemerons). In Sec.~\ref{sec:S3}, we return to the
previously found Minkowski solutions and establish a full set of
rules of the analytic continuation. Finally, in Sec.~\ref{sec:S4}, we
compute the Euclidean action of the ephemeron and show that its
topological charge is connected with the Thomas precession of its spin.
We conclude in Sec.~\ref{sec:S5}.

\section{The internal geometry of the ultrarelativistic nuclear collision}
\label{sec:S1}

The only tool which is capable of coping with the colored gauge fields in
the curved  geometry is the so-called tetrad formalism (see,
e.g.,\cite{Fock,Witten}). Indeed, the vector and spinor fields are
essentially defined in the tangent space.  In a tetrad basis, components
of any tensor (e.g  $A^{\alpha}(x)$, $\gamma^\alpha$) become scalars with
respect to a general coordinate transformations and behave like Lorentz
tensors under the local Lorentz group transformations.  The usual tensors
are then given by the tetrad decomposition,
$A^{\mu}(x)=e^{\mu}_{~\alpha}(x)A^\alpha(x)$,
$\gamma^{\mu}(x)=e^{\mu}_{~\alpha}(x)\gamma^\alpha$, etc. The covariant
derivative of the tetrad vector includes two connections (gauge fields).
One of them,  the Levi-Civita connection
$$\Gamma^{\lambda}_{~\mu\nu}={1\over 2} {\rm g}^{\lambda\rho}
\bigg[{\partial {\rm g}_{\rho\mu} \over \partial x^\nu } +{\partial {\rm
g}_{\rho\nu} \over \partial x^\mu } -{\partial {\rm g}_{\mu\nu} \over
\partial x^\rho } \bigg]~,$$ is the gauge  field which provides
covariance with respect to the general transformation of coordinates. The
second gauge field,  the spin connection
$\omega_{\mu}^{~\alpha\beta}(x)$, provides covariance with respect to the
local Lorentz rotation.

The curvilinear  metric of wedge dynamics in Minkowski space,
\begin{eqnarray}
ds_{\scriptscriptstyle M}^2=-d\tau_{\scriptscriptstyle M}^2+
\tau_{\scriptscriptstyle M}^2 d\eta_{\scriptscriptstyle M}^2 +dr^2+r^2
d\phi^2 ~, \label{eq:E1.1}\end{eqnarray} corresponds to the
parameterizations of the flat Minkowski space according to
\begin{eqnarray}
x^0=\pm\tau\cosh\eta~,~~~x^3=\pm\tau\sinh\eta~,\nonumber\\ x^1=
r\cos\phi~,~~~x^2=r\sin\phi~, \label{eq:E1.0}\end{eqnarray}
where $x^\mu =(\tau,r,\phi,\eta)$ are the contravariant components of the
curvilinear coordinates that cover the future (plus sign in
Eq.~(\ref{eq:E1.0}) ) and the past (minus sign) of the hyperplane
$t=0,~z=0$ of the interaction, and $x^\alpha
=(t,x,y,z)\equiv(x^0,x^1,x^2,x^3)$ are the Cartesian coordinates of the
Minkowski space.\footnote{We choose the polar coordinates in the
$xy$-plane, since we are interested in highly localized objects capable
of initiating a large $p_t$-transfer.} For the coordinates
(\ref{eq:E1.0}),  the four tetrad vectors $e^{\alpha}_{~\mu}$ form the
matrices
\begin{eqnarray}
e^{\alpha}_{~\mu}={\rm diag}(1,1,r,\tau)~,~~~
 e_{\alpha}^{~\mu}={\rm diag}(1,1,r^{-1},\tau^{-1})~.
\label{eq:E1.1a}\end{eqnarray} These vectors correctly reproduce the
curvilinear metric ${\rm g}_{\mu\nu}$ and the flat Minkowski metric
$g_{\alpha\beta}$, {\em i.e.},
\begin{eqnarray}
{\rm g}_{\mu\nu}=g_{\alpha\beta}e^{\alpha}_{~\mu}e^{\beta}_{~\nu} ={\rm
diag}[-1,1,r^2,\tau^2]~,\nonumber\\ g^{\alpha\beta}= {\rm g}^{\mu\nu}
e^{\alpha}_{~\mu}e^{\beta}_{~\nu}= {\rm diag}[-1,1,1,1]~.
\label{eq:E1.2}\end{eqnarray} The spin connection can be found from the
condition that the covariant derivative of the tetrad vectors is equal to
zero \cite{Witten},
\begin{eqnarray}
\nabla_\mu e^{a}_{~\nu}=\partial_\mu e^{a}_{~\nu}
+\omega^{~a}_{\mu~b}e^{b}_{~\nu} -
\Gamma^{\lambda}_{~\mu\nu}e^{a}_{~\lambda} =0~, \label{eq:E1.3a}
\end{eqnarray}
which can be solved for the components of $\omega$,
\begin{eqnarray}
\omega^{~\alpha\beta}_{\mu}=[\Gamma^{\lambda}_{~\mu\nu}
e^{\alpha}_{~\lambda}-
\partial_\mu e^{\alpha}_{~\nu}]e^{\beta~\nu}~.
\label{eq:E1.3}
\end{eqnarray}
Indeed, the tetrad vector $e^{\alpha}_{~\mu}$ is the coordinate vector
and the Lorentz vector at the same time. (The Lorentz index $\alpha$ and
the coordinate index $\mu$ are moved up and down by the local Minkowski
metric tensor $g_{\alpha\beta}$ and the global metric tensor ${\rm
g}_{\mu\nu}$, respectively.) The only non-vanishing components of the
connections are
\begin{eqnarray}
\Gamma^{\bf \cdot}_{\eta\eta\tau} = -\Gamma^{\bf
\cdot}_{\tau\eta\eta}=-\tau,~ \Gamma^{\bf \cdot}_{\phi\phi r} =
-\Gamma^{\bf  \cdot}_{r\phi\phi}=-r~,\nonumber\\
\omega_{\eta}^{~30}=-\omega_{\eta}^{~03}=1,~\omega_{\phi}^{~12}
=-\omega_{\phi}^{~21}=-1~. \label{eq:E1.4}\end{eqnarray}

The metric of the ``Euclidean'' counterpart of the  wedge dynamics is,
\begin{eqnarray}
ds_{\scriptscriptstyle E}^2=+d\tau_{\scriptscriptstyle E}^2
+\tau_{\scriptscriptstyle E}^2 d\eta_{\scriptscriptstyle E}^2 +dr^2+r^2
d\phi^2 ~. \label{eq:E1.5}\end{eqnarray}
In the tetrad formalism,  the transition to the Euclidean space is easily
done by making the time-like tetrad vector $e^{0}_{~\mu}$ imaginary,
$e^{0}_{~\mu}\to (e^{0}_{~\mu})_{\scriptscriptstyle E}=
(i,0,0,0),~~(e_{0}^{~\mu})_{\scriptscriptstyle E}= (-i,0,0,0)$. Then
Eqs.~(\ref{eq:E1.2}) take the form
\begin{eqnarray}
{\rm g}_{\mu\nu}= g_{\alpha\beta}(e^{\alpha}_{~\mu})_{\scriptscriptstyle
E} (e^{\beta}_{~\nu})_{\scriptscriptstyle E} ={\rm
diag}[1,1,r^2,\tau^2]~,\nonumber\\ g^{\alpha\beta}= {\rm g}^{\mu\nu}
(e^{\alpha}_{~\mu})_{\scriptscriptstyle E}
(e^{\beta}_{~\nu})_{\scriptscriptstyle E}= {\rm diag}[-1,1,1,1]~.
\label{eq:E1.6}\end{eqnarray} This formal step also leads to a set of
standard prescriptions for the transition to the Euclidean version of the
field theory, like $A^\tau_{\scriptscriptstyle E}=
(e_{0}^{~\tau})_{\scriptscriptstyle E}A^0 =-iA^0$ and
$\gamma^\tau_{\scriptscriptstyle E}= (e_{0}^{~\tau})_{\scriptscriptstyle
E}\gamma^0 =-i\gamma^0$. The same rule holds for the spin connection,
\begin{equation}
(\omega_{\mu}^{~03})_{\scriptscriptstyle M}~\to~
(\omega_{\mu}^{~03})_{\scriptscriptstyle E}= -i
(\omega_{\mu}^{~03})_{\scriptscriptstyle M}. \label{eq:E1.6a}
\end{equation}

These formulae indicate that we perform a transition to an {\em imaginary
proper time} $\tau$. Both metrics, (\ref{eq:E1.1}) and (\ref{eq:E1.5})
 are degenerate at $\tau=0$, because $|{\rm det}g_{\mu\nu}(x)|=
\tau^2 r^2$. The formal correspondence between these two metric forms is
given by $\tau_{\scriptscriptstyle M}=-i\tau_{\scriptscriptstyle E}$,
$~\eta_{\scriptscriptstyle M}=i\eta_{\scriptscriptstyle E}~.$

As we have discussed in the introduction, there is a remarkable
correspondence between the quantum veto on any measurements at the
intermediate stage of the exclusive process and the possibility to reduce
the underlying theory to the path-integral formalism in Euclidean space.
Being the solutions with the minimal action,  the instantons also provide
a local locking between the color and Euclidean spatial directions, thus
making the color virtually invisible. An important ingredient of this
structure is the $[O(4)]_{space}$ symmetry of an isotropic Euclidean space
which can be mapped onto the $[SU(2)\times SU(2)]_{color}$ group of the
color space. The presence of the spin connection explicitly provides the
theory with the local invariance with respect to $[O(4)]_{space}$
rotations. The reader can easily check that the spin connection of the
four-dimensional spherical coordinates exactly reproduces the pure-gauge
asymptote of the BPST instanton \cite{BPST}. The symmetry of the
Euclidean space with the metric (\ref{eq:E1.5}) does not allow for all
six rotations  of the  $[O(4)]_{space}$ group, since the spin connection
of this metric has only four (out of 12) non-vanishing components,
\begin{equation}\label{eq:E1.7}
\omega_{\eta}^{~03}=-\omega_{\eta}^{~30} =-1,~~~~~
\omega_{\phi}^{~12}=-\omega_{\phi}^{~21} =-1~.
\end{equation}
Hence, the system acquires an ``axis of quantization''. Since the spin
connection itself is the gauge field, it {\em can} be identified with the
pure gauge of the  Yang-Mills field. Then, in  the (iso-)vector
representation $A^{~\alpha\beta}_{\mu}$ of the gauge field of the $O(4)$
group, we must have
\begin{eqnarray}
[ A^{~\alpha\beta}_{\mu}]_{pure~gauge}=\omega^{~\alpha\beta}_{\mu}~,
\label{eq:E1.8}\end{eqnarray} and the color space also acquires a
quantization axis. The gauge fields of the $O(4)$ group have two
projections on its two $SU(2)$-subgroups,
\begin{eqnarray}
(A^{a}_{~\mu})_{\pm}={1\over 4}~
\eta^{a\alpha\beta}_{\pm}A^{~\alpha\beta}_{\mu}
 ={1\over 2}\big( \pm A^{0a}_{\mu} +
 {1\over 2}\epsilon^{a\alpha\beta}
 A^{~\alpha\beta}_{\mu}\big)~,
\label{eq:E1.9}\end{eqnarray}
where $\eta^{a\alpha\beta}_{\pm}$ are the 't Hooft symbols \cite{'t
Hooft}, and the subscripts $(\pm)$ denote two chiral
projections.\footnote{The 't Hooft symbols carry indices $\alpha\beta$ of
the local Cartesian coordinates, which in relativity theory correspond to
the local inertial coordinates. This parallel is more straightforward
that it might seem at the first glance. In the wedge dynamics, the local
inertial observers move with respect to each other with  velocities
that depend on the observers' coordinates.} Since $\eta^{a03}
=-\delta_{a3}$, and $\eta^{a12}=\delta_{a3}$, we have
\begin{eqnarray}
 (A^{3}_{~\eta})_{\pm}=\mp{1\over 2}~\omega^{~03}_{\eta}=\pm {1\over 2},~~
 (A^{3}_{~\phi})_{\pm}= {1\over 2}~\omega^{~12}_{\phi}= - {1\over 2},
\label{eq:E1.10}\end{eqnarray}
which is compatible with the gauge condition $A^\tau =0$  that we adopt
for both the Euclidean and the Lorentz regimes of the process. One can
easily find a representation for this potential which manifests its pure
gauge origin,
\begin{eqnarray}
A_{\mu}(x)=(1/2)A^{a}_{\mu}(x)\sigma^a=S\partial_\mu S^{-1}~.
\label{eq:E1.11}\end{eqnarray} Using the decomposition, $S=iu_0{\mathbf
1}+u_a{\mathbf\sigma}^a$, and $S^{-1}=-iu_0{\bf 1}+u_a {\mathbf
\sigma}^a$ we arrive at
\begin{eqnarray}
A_{\mu}(x)=(1/2)A^{c}_{\mu}(x){\bf\sigma}^c
=-(\epsilon^{abc}u_a\partial_\mu u_b + u_0\partial_\mu u_c-
u_c\partial_\mu u_0){\mathbf\sigma}^c~. \label{eq:E1.12}\end{eqnarray} By
comparison with (\ref{eq:E1.10}), and accounting for the unitarity,
$SS^{-1}=1$, we obtain a system of equations,
\begin{eqnarray}
 -4(u_1\partial_\eta u_2 -
u_2\partial_\eta u_1 + u_0\partial_\eta u_3 - u_3\partial_\eta u_0)=\pm
1~,\nonumber \\ -4(u_1\partial_\phi u_2 - u_2\partial_\phi u_1 +
u_0\partial_\phi u_3 - u_3\partial_\phi u_0)= - 1~,\nonumber\\ u_0^2 +
u_a^2=1~~, \label{eq:E1.13}
\end{eqnarray}
which has a solution
\begin{eqnarray}
(u_0)_{\pm}= \mp 2^{-1/2} \cos \eta/2,~~ (u_3)_{\pm}=  2^{-1/2} \sin
\eta/2~, \nonumber  \\ (u_1)_{\pm}=  2^{-1/2} \cos \phi/2,~~ (u_2)_{\pm}=
2^{-1/2} \sin \phi/2~. \label{eq:E1.14}\end{eqnarray} Thus, we can
conjecture that there exists a non-trivial solution of the Yang-Mills
equations which has a pure gauge asymptote
\begin{eqnarray}
(A^{3}_{~\eta})_{\pm} \to ~iS\partial_\eta S^{-1}= \pm {1\over
2},\nonumber  \\ (A^{3}_{~\phi})_{\pm} \to ~iS\partial_\phi S^{-1}=
-{1\over 2}. \label{eq:E1.15}\end{eqnarray}

This asymptote has only one component in color space. The $\pm$ signs
correspond to the right- and and left-handed projections on the
$SU(2)\times SU(2)$ group. The components $(A^{3}_{~\phi})_{\pm}$ are the
same for both projections, and  can be gauged out by one (Abelian) gauge
transformation. However, he components $(A^{3}_{~\eta})_{\pm}$ cannot.
Furthermore, in the wedge dynamics, the field $A_{\eta}$ is subjected to
a non-trivial boundary condition that completely fixes the gauge
$A^{\tau}=0$, while the component $A_{\phi}$ is free of this kind of
limitation.

\section{Self-dual solutions in the wedge dynamics.}
\label{sec:S2}

In this section, we find the solutions of Euclidean QED which are
compatible with the geometric background and boundary conditions of the
wedge dynamics. We need the solutions that provide the minimum of
Euclidean action. This minimum is well known to be reached on self-dual
and anti-self-dual solutions of the  Euclidean Yang-Mills equations. The
condition for the self-duality (anti-self-duality) of the field tensor
$F_{\mu\nu}$ reads as
\begin{eqnarray}
F^{\ast}_{\mu\lambda}\equiv {\rm g}_{\mu\nu} {\rm g}_{\lambda\sigma}
{\epsilon^{\nu\sigma\rho\kappa}\over 2\sqrt{\rm g}} F_{\rho\kappa} = \pm
F_{\mu\lambda}~. \label{eq:E2.1} \end{eqnarray}
Note that the definition of the dual tensor is different from the
familiar definition in flat space. This modification is obvious. Indeed,
the co- and contravariant tensor components are even of different
dimensions. We are looking for a self-dual solution, which delivers a
true minimum to the Euclidean action and becomes a pure gauge at
$\tau\to\infty$. This solution is supposed to approach the limiting value
given by the spin connection (\ref{eq:E1.15}) when $r\to 0$, and locally
in the rapidity direction (i.e., where the tetrad vectors
$(e^{\alpha}_{~\mu})_{\scriptscriptstyle E}$ form a {\em local Cartesian
basis}). Since the asymptote of the solution (after it is analytically
continued to Minkowski space, see Sec.~\ref{sec:S3}) has only one color
component at the Cauchy surface $\tau=const$, we shall look for a {\em
mono-colored} self-dual solution.\footnote{Such a solution will indeed be
found. Its stability with respect to the interaction with the fermion
modes or other gluon modes is a separate issue.} Embedded into the
$SU(3)$ color group, these solutions correspond to the gluon fields
$A^3_\mu$ and $A^8_\mu$, i.e., the so-called ``white gluons''.  Since the
Gell-Mann matrices $t^3_{ij}$ and $t^8_{ij}$ are diagonal, these gluons
(one in each of the three $SU(2)$ subgroups of the $SU(3)$ color group)
only differentiate quarks by their color, but they do not change the
color of a quark.

Let us choose the gauge of the wedge dynamics, $A_\tau=0$, and denote, $$
A^3_\phi =\Phi(\tau,r,\eta),~~~A^3_{\scriptscriptstyle
R}=R(\tau,r,\eta),~~~ A^3_\eta=N(\tau,r,\eta)~.$$ Since the field has
only one color component, the commutator in the definition,
$F_{\mu\nu}=\partial_\mu A_\nu - \partial_\nu A_\mu -[A_\mu ,A_\nu]$,
vanishes and the components of the field tensor are the same as in the
Abelian case where
\begin{eqnarray}
F_{\tau\eta}=\partial_\tau N,~~F_{\tau \phi}=\partial_\tau \Phi,~~
F_{\tau r}= \partial_\tau R ~, \nonumber \\ F_{r\eta}=\partial_r N
-\partial_\eta R,~~F_{r\phi}=\partial_r \Phi,~~
F_{\eta\phi}=\partial_\eta \Phi ~.
\label{eq:E2.0}\end{eqnarray}
The requirement of self-duality of the field (\ref{eq:E2.1}) yields a
system of equations,
\begin{eqnarray}
{\partial N \over\partial\tau}={\tau\over r}~{\partial\Phi\over\partial
r}~, \label{eq:E2.2} \end{eqnarray}
\begin{eqnarray}
{\partial \Phi \over \partial \tau} = - ~{r\over\tau}~\bigg[{\partial N
\over \partial r} - {\partial R \over \partial\eta }\bigg]~,
\label{eq:E2.3} \end{eqnarray}
\begin{eqnarray}
{\partial R\over\partial\tau}=-{1\over\tau
r}{\partial\Phi\over\partial\eta}~.
\label{eq:E2.4} \end{eqnarray}
The conditions of self-consistency for this system obviously coincide
with the Yang-Mills equations. For example, one of such conditions is $
\partial_\tau \partial_\eta R =\partial_\eta \partial_\tau R $. Using
Eqs.~(\ref{eq:E2.3}) and (\ref{eq:E2.4}), and excluding $N$ with the aid
of Eq.~(\ref{eq:E2.2}), we arrive at
\begin{eqnarray}
{\partial^2\Phi\over\partial\tau^2} +{1\over\tau}
{\partial\Phi\over\partial\tau}
+{1\over\tau^2}{\partial^2\Phi\over\partial\eta^2}=
-\bigg[{\partial^2\Phi\over\partial r^2}- {1\over
r}{\partial\Phi\over\partial r}\bigg]~. \label{eq:E2.5}\end{eqnarray}
This equation is easily solved by separation of variables. The solution
is readily found in the form,
\begin{eqnarray}
\Phi(\tau,r,\eta)= c ~ \lambda r ~ J_1(\lambda r)~ K_{-i\nu}(\lambda
\tau) e^{-\nu\eta} \label{eq:E2.6}\end{eqnarray} In order to find similar
equations for $R$ and $N$, we have to use the self-consistency conditions
of the third order. To derive them, we must begin with three equations of
the second order,  $\partial_\tau\partial_r \Phi =\partial_r
\partial_\tau \Phi,~~~ \partial_\eta \partial_r \Phi =\partial_r
\partial_\eta \Phi,~~~ \partial_\eta \partial_\tau \Phi =\partial_\tau
\partial_\eta \Phi,$ which read as
\begin{eqnarray}
{\partial\over\partial\tau} \bigg({r\over\tau}{\partial
N\over\partial\tau}\bigg) ={\partial\over\partial r}\bigg(
{r\over\tau}{\partial R\over\partial\eta} -{r\over\tau}{\partial
N\over\partial r}\bigg)~, \label{eq:E2.7}\end{eqnarray}
\begin{eqnarray}
{\partial\over\partial\eta} \bigg({r\over\tau}{\partial
N\over\partial\tau}\bigg) =-{\partial\over\partial r} \bigg( \tau r
{\partial R\over\partial\tau}\bigg)~, \label{eq:E2.8}\end{eqnarray}
\begin{eqnarray}
{\partial\over\partial\eta}\bigg({r\over\tau}{\partial N\over\partial r}
-{r\over\tau}{\partial R\over\partial\eta}\bigg)
={\partial\over\partial\tau} \bigg( \tau r {\partial
R\over\partial\tau}\bigg)~.
\label{eq:E2.9} \end{eqnarray}
Excluding $N$ from Eqs.~(\ref{eq:E2.8}) and (\ref{eq:E2.9}), we derive
an equation for $R$,
\begin{eqnarray}
\bigg[ {\partial^2\over\partial\tau^2}+ {1\over\tau}
{\partial\over\partial\tau}
+{1\over\tau^2}{\partial^2\over\partial_\eta^2}\bigg] \bigg(\tau{\partial
R\over\partial\tau}\bigg) =-\bigg[ {\partial^2\over\partial r^2}+ {1\over
r}{\partial\over\partial r} -{1\over r^2}\bigg] \bigg(\tau{\partial
R\over\partial\tau}\bigg)~,
\label{eq:E2.10} \end{eqnarray}
with the solution,
\begin{eqnarray}
\tau{\partial R\over\partial\tau}= c_1~\lambda_1 J_1(\lambda_1 r)
K_{-i\nu_1}(\lambda_1 \tau) e^{-\nu_1 \eta}~.
\label{eq:E2.11}
\end{eqnarray}
In the same way, excluding $R$ from Eqs.~(\ref{eq:E2.7})
and (\ref{eq:E2.8}), we derive an equation for $N$,
\begin{eqnarray}
\bigg[ {\partial^2\over\partial\tau^2}+ {1\over\tau}
{\partial\over\partial\tau}
+{1\over\tau^2}{\partial^2\over\partial_\eta^2}\bigg]
\bigg({1\over\tau}{\partial N\over\partial\tau}\bigg) =-\bigg[
{\partial^2\over\partial r^2}+{1\over r}{\partial\over\partial r}\bigg]
\bigg({1\over\tau}{\partial N \over\partial\tau}\bigg)~, \label{eq:E2.12}
\end{eqnarray} which has the solution,
\begin{eqnarray}
{1\over\tau}~{\partial N\over\partial\tau}= c_2~ J_0(\lambda_2 r)
K_{-i\nu_2}(\lambda_2 \tau) e^{-\nu_2 \eta}~.
\label{eq:E2.13}
\end{eqnarray}
Substituting the solutions (\ref{eq:E2.6}), (\ref{eq:E2.11}), and
(\ref{eq:E2.13}) into the original system
(\ref{eq:E2.2})-(\ref{eq:E2.4}), we find the following relations among
the constants, $$\nu_1=\nu_2 =\nu,~~~ \lambda_1= \lambda_2=\lambda,~~~
c_2=\pm c,~~~c_1= \mp\nu c~.$$ The second solutions of ordinary
differential equations that we solve after the separation of variables
are dropped for various reasons: The Bessel function
$I_{-i\nu}(\lambda\tau)$ explodes at large $\tau$. The Bessel functions
$Y_0(\lambda r)$ and $Y_1(\lambda r)$ are singular at the origin. The
second exponent, $e^{+\nu\eta}$ is accounted for, since
$K_\nu(x)=K_{-\nu}(x)$. Finally, we may write the solution as
\begin{eqnarray}
\Phi_{\nu,\lambda,\vec{r}_0}(x)=c \lambda r J_1(\lambda
r)K_{-i\nu}(\lambda\tau) e^{-\nu\eta}~, \nonumber \\
R_{\nu,\lambda,\vec{r}_0}(x)= \mp c~\nu \lambda  J_1(\lambda r)
M_{-1,-i\nu}(\lambda\tau) e^{-\nu\eta}~, \nonumber \\
N_{\nu,\lambda,\vec{r}_0}(x)= \pm c~ J_0(\lambda r)
M_{1,-i\nu}(\lambda\tau) e^{-\nu\eta}~, \label{eq:E2.14} \end{eqnarray}
where $r=|\vec{r}-\vec{r}_0|$, and $\vec{r}_0$ labels the position of the
center in the transverse $xy$-plane. The new functions,
\begin{eqnarray}
M_{m,-i\nu}(\tau)=\int_\infty^\tau K_{-i\nu}(t)~t^m ~dt,
\label{eq:E2.15}
\end{eqnarray}
are similar to the functions $R^{(j)}_{m,-i\nu}(\tau)=\int
H^{(j)}_{-i\nu}(t)~t^m~dt$, used extensively in paper [III]. A particular
choice of the infinite lower limit in Eq.~(\ref{eq:E2.15}) leads to the
solutions that vanish at $\tau\to\infty$. Another form of the solution,
in which the variables are not separated explicitly, but are extremely
useful for the future analysis, is obtained by means of the (two-sided)
Laplace transform, e.g.,
\begin{eqnarray}
\Phi_{\theta,\lambda,\vec{r}_0}(x)= \int_{-\infty}^{\infty} e^{\nu\eta}
\Phi_{\nu,\lambda,\vec{r}_0}(x) d\nu~,
\label{eq:E2.16} \end{eqnarray} In
this way, we trade the ``quantum number'' $\nu$ for the quantum number
$\theta$ which is the Euclidean analog of the Minkowski rapidity.
Computing the Laplace integral, we use the formula,
\begin{eqnarray}
\int_{-\infty}^{\infty} e^{\nu(\theta-\eta)}~K_{i\nu}(\lambda\tau)~d\nu=
2\int_{0}^{\infty} \cosh[\nu(\eta-\theta)]~K_{i\nu}(\lambda\tau)~d\nu =
\pi e^{-\lambda\tau\cos(\eta -\theta)}~,
\label{eq:E2.17} \end{eqnarray}
where the integral is convergent at $|{\rm Re}(\eta-\theta)|\leq\pi/2$,
and can be analytically continued. The solution with
a given Euclidean rapidity $\theta$ reads as
\begin{eqnarray}
\Phi_{ \theta,\lambda,\vec{r}_0}(x_{\scriptscriptstyle E})= c~\lambda r
J_1(\lambda r) e^{-\lambda\tau_{\scriptscriptstyle E}
\cos(\eta_{\scriptscriptstyle E}- \theta_{\scriptscriptstyle E})}~,
\label{eq:E2.18a} \end{eqnarray}
\begin{eqnarray}
R_{\theta,\lambda,\vec{r}_0}(x_{\scriptscriptstyle E})= \mp c ~\lambda
J_1(\lambda r) \tan(\eta_{\scriptscriptstyle
E}-\theta_{\scriptscriptstyle E})
 e^{-\lambda\tau_{\scriptscriptstyle E}
 \cos(\eta_{\scriptscriptstyle E}-
 \theta_{\scriptscriptstyle E})}~, \nonumber \\
N_{\lambda,\vec{r}_0}(x_{\scriptscriptstyle E})=\mp c~ J_0(\lambda r)~
{1+\lambda\tau_{\scriptscriptstyle E} \cos(\eta_{\scriptscriptstyle
E}-\theta_{\scriptscriptstyle E})\over \cos^2(\eta_{\scriptscriptstyle
E}-\theta_{\scriptscriptstyle E})} e^{-\lambda\tau_{\scriptscriptstyle E}
\cos(\eta_{\scriptscriptstyle E}-\theta_{\scriptscriptstyle E})}~,
\label{eq:E2.18} \end{eqnarray}
where we have explicitly labeled the Euclidean variables. The condition
$|\eta_{\scriptscriptstyle E }-\theta_{\scriptscriptstyle E}|\leq\pi/2$
means that these solutions occupy only a quarter of the Euclidean
$t_{\scriptscriptstyle E}z_{\scriptscriptstyle E}$-plane. Indeed, in
order to satisfy this condition, we must have
$|\eta_{\scriptscriptstyle E}| \leq\pi/4$ and
$|\theta_{\scriptscriptstyle E}|\leq\pi/4$, satisfied separately.
Introducing the Euclidean energy $\omega_{\scriptscriptstyle E}=
\lambda\cos\theta$ and the Euclidean longitudinal momentum
$p^z_{\scriptscriptstyle E}=\lambda\sin\theta$, along with
$t_{\scriptscriptstyle E}= \tau_{\scriptscriptstyle E}
\cos\eta_{\scriptscriptstyle E}$ and $z_{\scriptscriptstyle E}=
\tau_{\scriptscriptstyle E}\sin\eta_{\scriptscriptstyle E}$, we may
rewrite the exponent in Eqs.~(\ref{eq:E2.18}) as
$$e^{-\lambda\tau_{\scriptscriptstyle E}\cos(\eta_{\scriptscriptstyle
E}-\theta_{\scriptscriptstyle E})}= e^{-(\omega_{\scriptscriptstyle
E}t_{\scriptscriptstyle E}+ p^z_{\scriptscriptstyle E}
z_{\scriptscriptstyle E})},$$ which clearly indicates that our solutions
are inhomogeneous waves adjacent  to the hypersurface
$\tau_{\scriptscriptstyle E}^2=t_{\scriptscriptstyle E}^2
+z_{\scriptscriptstyle E}^2=0$. Another form of the solution can be
obtained by means of the (Abelian) gauge transform, i.e., by adding the
gradient, $$A'_\mu(x_{\scriptscriptstyle E}) =
A_\mu(x_{\scriptscriptstyle E}) -
\partial_\mu \chi(x_{\scriptscriptstyle E}), $$
of the function
\begin{eqnarray}
\chi(x_{\scriptscriptstyle E}) = \mp c~\tan(\eta_{\scriptscriptstyle
E}-\theta_{\scriptscriptstyle E}) J_0(\lambda r)~.
\label{eq:E2.19}
\end{eqnarray}
This transform does not modify $A_\phi=\Phi$, and it results in the field
that satisfies the boundary condition,
$A_\eta(\tau_{\scriptscriptstyle E}=0)=0$,
\begin{eqnarray}
R'_{\theta,\lambda,\vec{r}_0}(x_{\scriptscriptstyle E})= \mp c \lambda
J_1(\lambda r) \tan(\eta_{\scriptscriptstyle
E}-\theta_{\scriptscriptstyle E}) [e^{-\lambda\tau_{\scriptscriptstyle E}
\cos(\eta_{\scriptscriptstyle E}-\theta_{\scriptscriptstyle E})}-1]~,
\nonumber \\ N'_{\lambda,\vec{r}_0}(x_{\scriptscriptstyle E})= \mp c
J_0(\lambda r) \bigg[{e^{-\lambda\tau_{_{\scriptscriptstyle E}}
\cos(\eta_{_{\scriptscriptstyle E}}-\theta_{_{\scriptscriptstyle
E}})}-1\over \cos^2(\eta_{\scriptscriptstyle
E}-\theta_{\scriptscriptstyle E})} +\lambda\tau_{\scriptscriptstyle E}
{e^{-\lambda\tau_{\scriptscriptstyle E} \cos(\eta_{\scriptscriptstyle
E}-\theta_{\scriptscriptstyle E})}\over \cos(\eta_{\scriptscriptstyle
E}-\theta_{\scriptscriptstyle E})}\bigg]~.
\label{eq:E2.20} \end{eqnarray}
When $\tau\to\infty$, the solution with  components (\ref{eq:E2.18a})
and (\ref{eq:E2.20}) becomes a pure gauge; then it is just the gradient
of the function (\ref{eq:E2.19}). Furthermore, along a singular line
$r=0$ and at $\eta_{\scriptscriptstyle E}= \theta_{\scriptscriptstyle E}$
this pure gauge potential approaches its asymptotic value which coincides
with the projection of the $O(4)$ spin connection onto the $SU(2$)
potentials. Since at large $\tau$, the states of the wedge dynamics are
highly localized in the rapidity direction around the rapidity $\theta$,
it is quite natural that the limit of the spin connection is reached
exactly at $\eta=\theta$. In order to comply with the asymptotic
condition (\ref{eq:E1.15}), we have to take $c=1/2$, though in general,
we have the linear equations which leave the common normalization factor
not defined.

An important remark should be made at this point. {\em The pure-gauge
asymptotic behavior of this solution is enforced by the boundary
condition imposed on the potential $A_\eta$ at $\tau=0$.} This is a
physical requirement of continuity brought from the physical Minkowski
world.

The pure-gauge asymptote of Eq.~(\ref{eq:E2.20}) becomes singular at the
two isolated points, $\theta_{\scriptscriptstyle E} =\pi/4$,
$\eta_{\scriptscriptstyle E}= -\pi/4$, and $\theta_{\scriptscriptstyle E}
=-\pi/4$, $\eta_{\scriptscriptstyle E}= \pi/4$, which correspond to the
limit of the infinite momentum frame (or the null-plane dynamics). In the
same limit the solution (\ref{eq:E2.18}) (which otherwise tends to zero
at $\tau\to\infty$), is also singular. This solution does not obey the
boundary conditions that provide a complete fixing of the gauge
$A^\tau=0$.  The solution (\ref{eq:E2.20}) is regular in all these
limits.\footnote{These limits are sensitive to the order in which they
are applied. This ambiguity can be traced back to the Minkowski space,
where the normal and tangent direction on the hypersurface of the
constant $\tau\to 0$ are degenerate. The way one should resolve this
ambiguity depends on the  observable under consideration.}
 For large values of the  parameter $\lambda$, which can be put in
correspondence with the transverse momentum $p_t$ of the plane-wave
solutions in Minkowski space, the Euclidean life-time of the solution
given by Eqs.~(\ref{eq:E2.18a}) and (\ref{eq:E2.20}) is very
short.\footnote{  We would suggest to call this object {\em ephemeron} in
order not to create an image of a particle and emphasize the ephemeral
nature of this field configuration.} It has a noticeable amplitude only in the
vicinity of $\tau_{\scriptscriptstyle E}=0$ and thus, it very much resembles
surface (Tamm) states in condensed matter physics.
In the next section we show, that the ephemeron solutions can be
analytically continued to Minkowski space, where they immediately become
the asymptotic states, similar to those truncated by the LSZ reduction
procedure in the standard scattering problem.

\section{Propagating states of wedge dynamics.}
\label{sec:S3}

In this section we show that the ephemeron solutions in Euclidean space
can be obtained by the analytic continuation of the propagating waves of
the wedge dynamics. This continuation involves not only the proper time,
but the rapidity variables also.\footnote{Recently, Shuryak and Zahed
used the analytic continuation to Euclidean rapidity in their attempt to
estimate the effect of instantons on the high-energy exclusive scattering
between two {\em a priori} factorized  partons which are treated in the
eikonal approximation \cite{ShZa}.} In the framework of wedge dynamics
with the gauge $A_\tau =0$, the solutions of the linearized Yang-Mills
equations (thus, with a given color, which is not indicated) were found
in paper [III]. The two modes of the radiation field with components
$A_m=(A_x,A_y,A_\eta)$ are
\begin{eqnarray}
V^{(TE)}_{{\vec k},\nu}(x)={e^{-\pi\nu/2}\over 2^{5/2}\pi k_{t}} \left(
\begin{array}{c} k_y \\ -k_x \\ 0 \end{array} \right) H^{(2)}_{-i\nu}
(k_{t}\tau) e^{i\nu\eta +i{\vec k}{\vec r}}~,\nonumber\\
% {\rm and}~~~
V^{(TM)}_{{\vec k},\nu}(x)={e^{-\pi\nu/2}\over 2^{5/2}\pi k_{t}} \left(
\begin{array}{c}
                 \nu k_x R^{(2)}_{-1,-i\nu}(k_{t}\tau) \\
                 \nu k_y R^{(2)}_{-1,-i\nu}(k_{t}\tau) \\
                 - R^{(2)}_{1,-i\nu}(k_{t}\tau)
                                            \end{array} \right)
                e^{i\nu\eta +i{\vec k}{\vec r}}~.
\label{eq:E3.1}\end{eqnarray} The mode $V^{(TM)}$ is constructed from the
functions  (see paper [III]) $R^{(2)}_{m,-i\nu}(x)=\int
H^{(2)}_{-i\nu}(t)t^m dt$ (originally defined as the indefinite
integrals) corresponding to the boundary condition of vanishing gauge
field at $\tau=0$. This guarantees continuous behavior of the field at
$\tau=0$. Indeed, as $\tau\rightarrow 0$,~ the normal and the tangential
directions become degenerate. As long as $A^\tau=0$ is the gauge
condition,  continuity requires that  $A^\eta\rightarrow 0$ as
$\tau\rightarrow 0$. The transverse electric mode $V^{(TE)}$ and
transverse magnetic mode $V^{(TM)}$ have quantum numbers of boost $\nu$
and of transverse momentum $\vec{k}=(k_x,k_y)$. Let us trade the
latter for the position $\vec{r}_0$ of the azimuthally symmetric mode
$V_{\nu,\lambda\vec{r}_0}$, acting as follows. We introduce
\begin{eqnarray}
V^m_{\vec{r}_0,\nu}(x)=\int~{d^2\vec{k}\over 2\pi}~ V^m_{{\vec k},\nu}(x)~ 
e^{- i{\vec k}{\vec r_0}}~, \label{eq:E3.2} \end{eqnarray} 
and integrate only
over the angle between the vectors $\vec{k}$ and ${\vec r} -{\vec r_0}$,
leaving the $dk_t$ integral not integrated, and
denoting (for convenience) $k_t$ as $\lambda$.
Furthermore, let us write the answer in terms of the radial and azimuthal
components $A_{r}$ and $A_\phi$,
\begin{eqnarray}
A_r=A^r=\cos\phi ~A^x + \sin\phi ~A^y, ~~~ A_\phi =r^2A^\phi =-r\sin\phi
~A^x + r\cos\phi ~A^y ~.
\label{eq:E3.3} \end{eqnarray}
The transverse electric mode $V^{(TE)}$ has only a $\phi$-component,
\begin{eqnarray}
\Phi_{\nu,\lambda,\vec{r}_0}(x_{\scriptscriptstyle M})=-2^{-5/2}
\pi^{-1} \lambda
r J_1(\lambda r) e^{-\pi\nu/2} H^{(2)}_{-i\nu}(\lambda\tau)e^{i\nu\eta}~,
\label{eq:E3.4} \end{eqnarray}
The transverse magnetic  mode $V^{(TM)}$ has only $r$- and
$\eta$-components,
\begin{eqnarray}
R_{\nu,\lambda,\vec{r}_0}(x_{\scriptscriptstyle M})=2^{-5/2}
\pi^{-1}\nu \lambda J_1(\lambda r) e^{-\pi\nu/2}
R^{(2)}_{-1,-i\nu}(\lambda\tau)e^{i\nu\eta}~, \label{eq:E3.5}
\end{eqnarray}
\begin{eqnarray}
N_{\nu,\lambda,\vec{r}_0}(x_{\scriptscriptstyle M})=-2^{-5/2} \pi^{-1} i
J_0(\lambda r) e^{-\pi\nu/2} R^{(2)}_{1,-i\nu}(\lambda\tau)e^{i\nu\eta}~,
\label{eq:E3.6} \end{eqnarray}
where, as previously, $r=|\vec{r}-\vec{r}_0|$. Now, one can immediately
establish the connection between the components
(\ref{eq:E3.4})-(\ref{eq:E3.6}) of the radiation field in Minkowski space
and the components (\ref{eq:E2.14}) of the self-dual solution in
Euclidean space. Since
\begin{eqnarray}
-{i\pi\over 2}~e^{-\pi\nu/2}
H^{(2)}_{-i\nu}(\lambda\tau_{\scriptscriptstyle M})= -{i\pi\over
2}~e^{-\pi\nu/2} H^{(2)}_{-i\nu}(\lambda\tau_{\scriptscriptstyle E}
e^{-i\pi/2})= K_{-i\nu}(\lambda\tau_{\scriptscriptstyle E}) ~,
\label{eq:E3.7} \end{eqnarray}
we may guess that the desired prescription is,
$$\tau_{\scriptscriptstyle M} = e^{-i\pi/2}\tau_{\scriptscriptstyle E}~,
\eta_{\scriptscriptstyle M} = e^{i\pi/2}\eta_{\scriptscriptstyle E}~.$$
Since the function $K_{i\nu}(z)$ has a branching point at the origin,
this conjecture is not yet completely certain. The issue can be clarified
only after we get the Minkowski modes with the quantum number of
rapidity. These modes can be obtained by means of the Fourier transform,

\begin{eqnarray}
V_{\theta,\lambda,\vec{r}_0 }(x) = \int_{-\infty}^{+\infty}  {d\nu \over
(2\pi)^{1/2}} e^{-i\nu\theta} V_{\nu,\lambda,\vec{r}_0}(x)~.
\label{eq:E3.9}
\end{eqnarray}
Using the following integral representation for the Hankel functions,
\begin{eqnarray}
e^{-\pi\nu/2}e^{i\nu\eta} H^{(2)}_{- i\nu} (\lambda\tau)= {i\over \pi}
\int_{-\infty}^{\infty} e^{- i\lambda\tau \cosh (\theta-\eta)}
e^{i\nu\theta} d \theta~~,
\label{eq:E3.10}
\end{eqnarray}
and changing the order of integration in the expressions (\ref{eq:E3.5})
and (\ref{eq:E3.6}) for the $A_r$ and $A_\eta$ components, we arrive at
\begin{eqnarray}
\Phi_{\theta,\lambda,\vec{r}_0}(x_{\scriptscriptstyle M})= {1\over
4\pi^{3/2}} \lambda r J_1(\lambda r) e^{-
i\lambda\tau_{\scriptscriptstyle M} \cosh (\theta_{\scriptscriptstyle
M}-\eta_{\scriptscriptstyle M})}~,
\label{eq:E3.11} \end{eqnarray}
\begin{eqnarray}
R'_{\theta,\lambda,\vec{r}_0}(x_{\scriptscriptstyle M})= {i\over
4\pi^{3/2}} \lambda  J_1(\lambda r) \tanh(\eta_{\scriptscriptstyle
M}-\theta_{\scriptscriptstyle M}) \big[e^{-
i\lambda\tau_{\scriptscriptstyle M} \cosh (\theta_{\scriptscriptstyle
M}-\eta_{\scriptscriptstyle M})} -1 \big]~,
\label{eq:E3.12}
\end{eqnarray}
\begin{eqnarray}
N'_{\theta,\lambda,\vec{r}_0}(x_{\scriptscriptstyle M}) = - {i\over
4\pi^{3/2}}
 J_0(\lambda r)
\bigg[ {e^{-i\lambda\tau_{\scriptscriptstyle M}
\cosh(\theta_{\scriptscriptstyle M}-\eta_{\scriptscriptstyle M})} -1
\over \cosh^2(\theta_{\scriptscriptstyle M}-\eta_{\scriptscriptstyle M})}
+i\lambda\tau_{\scriptscriptstyle M}
{e^{-i\lambda\tau_{\scriptscriptstyle M} \cosh(\theta_{\scriptscriptstyle
M}-\eta_{\scriptscriptstyle M})} \over \cosh(\theta_{\scriptscriptstyle
M}-\eta_{\scriptscriptstyle M})} \bigg]~.
\label{eq:E3.13} \end{eqnarray}
This representation clearly satisfies the boundary condition,
$A_\eta(\tau_{\scriptscriptstyle M}=0)=0$. As it has been discussed in
paper [III], this boundary condition fixes the gauge $A^\tau=0$
completely. The solution that does not satisfy this boundary condition,
but can be ``analytically continued'' to the solution (\ref{eq:E2.20}) in
Euclidean space can be obtained by means of the gauge transform,
$$A_\mu(x_{\scriptscriptstyle M}) =A'_\mu(x_{\scriptscriptstyle
M})+\partial_\mu \chi(x_{\scriptscriptstyle M}), $$ with the function
\begin{eqnarray}
\chi(x_{\scriptscriptstyle M}) = i~\tanh(\eta_{\scriptscriptstyle
M}-\theta_{\scriptscriptstyle M}) J_0(\lambda r)~, \label{eq:E3.14}
\end{eqnarray} and is as follows,
\begin{eqnarray}
R_{\theta,\lambda,\vec{r}_0}(x_{\scriptscriptstyle M})= {i\over
4\pi^{3/2}} \lambda  J_1(\lambda r) \tanh(\eta_{\scriptscriptstyle
M}-\theta_{\scriptscriptstyle M}) e^{- i\lambda\tau_{\scriptscriptstyle
M} \cosh (\theta_{\scriptscriptstyle M}-\eta_{\scriptscriptstyle M})}~,
\label{eq:E3.15} \end{eqnarray}
\begin{eqnarray}
N_{\nu,\lambda,\vec{r}_0}(x_{\scriptscriptstyle M}) = - {i\over
4\pi^{3/2}}
 J_0(\lambda r)
\bigg[ {e^{-i\lambda\tau_{\scriptscriptstyle M}
\cosh(\theta_{\scriptscriptstyle M}-\eta_{\scriptscriptstyle M})}  \over
\cosh^2(\theta_{\scriptscriptstyle M}-\eta_{\scriptscriptstyle M})}
+i\lambda\tau_{\scriptscriptstyle M}
{e^{-i\lambda\tau_{\scriptscriptstyle M} \cosh(\theta_{\scriptscriptstyle
M}-\eta_{\scriptscriptstyle M})} \over 
\cosh(\theta_{\scriptscriptstyle M}-\eta_{\scriptscriptstyle M})} \bigg]~. 
\label{eq:E3.16} \end{eqnarray}
Comparing Eqs.~(\ref{eq:E3.11})-(\ref{eq:E3.13}) with the expressions
(\ref{eq:E2.18a}) and (\ref{eq:E2.20}) of the Euclidean solutions, we can
formulate the complete set of substitutions that are required to pass
over from the Minkowski solutions, to the Euclidean ones,
\begin{eqnarray}
\tau_{\scriptscriptstyle M}\to -i\tau_{\scriptscriptstyle E},~~~
\eta_{\scriptscriptstyle M}\to i\eta_{\scriptscriptstyle E},~~~
\theta_{\scriptscriptstyle M}\to 
i\theta_{\scriptscriptstyle E}~,\nonumber\\ 
|\eta_{\scriptscriptstyle E}|\leq\pi/4~,~~~
|\theta_{\scriptscriptstyle E}|\leq\pi/4~,\nonumber\\
A_\phi(x_{\scriptscriptstyle M})\to 
A_\phi(-i\tau_{\scriptscriptstyle E},i\eta_{\scriptscriptstyle E})=
 A_\phi(x_{\scriptscriptstyle E})~,\nonumber\\ 
A_r(x_{\scriptscriptstyle M})\to
A_r(-i\tau_{\scriptscriptstyle E},i\eta_{\scriptscriptstyle E})=
A_r(x_{\scriptscriptstyle E})~,\nonumber\\ 
A_\eta(x_{\scriptscriptstyle M})\to 
-i~A_\eta(-i\tau_{\scriptscriptstyle E},i\eta_{\scriptscriptstyle E})= 
A_\eta(x_{\scriptscriptstyle E})~. 
\label{eq:E3.17} \end{eqnarray}
On can easily see that the two signs of $A_r$ and $A_\eta$ of the
Euclidean solutions correspond to the two linear combinations,
$V^{\pm}=2^{-1/2}(V^{(TE)}\pm V^{(TM)})$, which are
the propagating waves with two circular
polarizations\footnote{ The Minkowski solutions $V^{\pm}$ have an extra factor
$(2\pi)^{-3/2}$ due to the physical normalization of these solutions as
the one-particle states.}. The last one of these equations duplicates
Eq.(\ref{eq:E1.6a}) and thus, is consistent with the view of the
transition to the Euclidean metric as an analytic continuation of the
time-like tetrad vector $e_{0}^{~\tau}$. If we assume that the final
states occupy the entire future domain of the collision, then this
transition takes place on the hypersurface $\tau=0$ where the metric of
the wedge dynamics (both in Minkowski and Euclidean versions) is
degenerate, i.e., ${\rm det}|g_{\mu\nu}|=0$.

The analytic continuation of the tetrad vector $e_{0}^{~\tau}$ can be
employed to find the Euclidean version of the fermion wave function. In
Minkowski space,  the positive-frequency right-handed state is described
by
\begin{eqnarray}
\psi_{R}(p_t,\theta_{\scriptscriptstyle M}; \tau_{\scriptscriptstyle
M},\eta_{\scriptscriptstyle M})= {1\over 4\pi^{3/2}p_t}~\left(
\begin{array}{c} e^{(\eta_{\scriptscriptstyle
M}-\theta_{\scriptscriptstyle M})/2}~p_t \\
 -e^{-(\eta_{\scriptscriptstyle M}-\theta_{\scriptscriptstyle M})/2}~
 (p_x+ip_y) \end{array} \right)
           ~e^{-i p_t\tau_{\scriptscriptstyle M}
 \cosh(\eta_{\scriptscriptstyle M}-\theta_{\scriptscriptstyle M})} ~.
\label{eq:E3.18}\end{eqnarray}
After analytic continuation, this spinor becomes
\begin{eqnarray}
\psi_{R}(p_t,\theta_{\scriptscriptstyle E}; \tau_{\scriptscriptstyle
E},\eta_{\scriptscriptstyle E})= {1\over 4\pi^{3/2}p_t}~\left(
\begin{array}{c} e^{i(\eta_{\scriptscriptstyle
E}-\theta_{\scriptscriptstyle E})/2}~p_t \\
 -e^{-i(\eta_{\scriptscriptstyle E}-\theta_{\scriptscriptstyle E})/2}~
 (p_x+ip_y) \end{array} \right)
           ~e^{- p_t \tau_{\scriptscriptstyle E}
 \cos(\eta_{\scriptscriptstyle E}-\theta_{\scriptscriptstyle E})} ~,
\label{eq:E3.19}\end{eqnarray} which also corresponds to the
inhomogeneous wave that flares up for the time $\sim 1/p_t$ near the edge
($\tau=0$)  of the Euclidean evolution.

\section{Euclidean action and topological charge.}
\label{sec:S4}

As  has been explained in the Introduction, we are looking for the
fluctuations that can minimize the Euclidean action of the system of
colliding nuclei along the path that ends at the moment of the nuclei
intersection. One may hope that this optimal path will also include the
ephemeron configurations which can be then continued to Minkowski space
as the propagating modes of the expanding quark-gluon system. Starting
from the fields given by the Eqs.~(\ref{eq:E2.0}) we may find the
Euclidean action of the ephemeron,
\begin{eqnarray}
 S_{\scriptscriptstyle E} =
 {1\over 4 g^2}\int d^4x \sqrt{\rm g}~ {\rm g}^{\mu\rho}
 {\rm g}^{\nu\sigma} F^a_{\mu\nu} F^a_{\rho\sigma} \nonumber\\
 ={1\over  g^2} \int d^4x ~\tau r~
 \bigg[{1\over\tau^2}\bigg({\partial\Phi\over\partial\tau}\bigg)^2+
 {1\over r^2}\bigg({\partial N\over\partial\tau}\bigg)^2
 + \bigg({\partial R \over\partial\tau}\bigg)^2\bigg]~.
\label{eq:E4.1}\end{eqnarray}

In the same way, we compute the topological charge
\begin{eqnarray}
 Q= {1\over 32 \pi^2}\int d^4x \sqrt{\rm g}
 ~{\epsilon^{\mu\nu\rho\sigma}\over 2\sqrt{\rm g}}
 ~ F^a_{\mu\nu} F^a_{\rho\sigma}=
 \pm~{ g^2\over 8\pi^2}~S_{\scriptscriptstyle E} ~.
 \label{eq:E4.3}\end{eqnarray}
Thus, we have a standard relation between the  action and the winding
number, well known for the instanton field configurations, which is true
for any self-dual Euclidean field. Using the expressions
(\ref{eq:E2.18a}) and (\ref{eq:E2.20}) for the field components, we
arrive at
\begin{eqnarray}
 S_{\scriptscriptstyle E}
 ={c^2\lambda^4 \over  g^2} \int_{0}^{\infty} \tau d\tau
 \int_{0}^{R_t}r d r~\int_{0}^{2\pi}d\phi\int_{-\pi/4}^{\pi/4}d\eta
~e^{-2\lambda\tau \cos(\eta-\theta)}
 \big[J_0^2(\lambda r)+ J_1^2(\lambda r)\big]~,
\label{eq:E4.4}\end{eqnarray} 
where the upper limit $R_t$ in the integral
over the transverse radius $r$ is introduced in order to isolate the
divergence naturally inherent in the action of the free infinitely
propagating field. The value of this parameter is limited from above
by the transverse area of the nuclei intersection. Practically, the upper
limit is much lower, since the ephemeron with large $\lambda$ can be
resolved only by a process with the large transverse momentum transfer.
Thus, $R_t$ should be of the order of $\lambda^{-1}$. All integrations in
Eq.~(\ref{eq:E4.4}) can be carried out analytically and yield
\begin{eqnarray}
 S_{\scriptscriptstyle E} =
 {c^2\pi\lambda^2 R_t^2 \over  g^2}~~
 {1 \over\cos 2\theta_{\scriptscriptstyle E}}~
 \bigg[J_0^2(\lambda R_t)+ J_1^2(\lambda R_t)
 - {J_0(\lambda R_t) J_1(\lambda R_t)\over \lambda R_t}\bigg]~.
\label{eq:E4.5}\end{eqnarray} 
Even at finite $R_t$, this action diverges
at $\theta \to \pm\pi/4$, i.e., when the Minkowski rapidity of the
particle tends to infinity and thus it  infinitely moves freely (in
Minkowski space). When  $\lambda R_t > 1$, the asymptotic expansion of
the Bessel functions is sufficiently accurate. Then
\begin{eqnarray}
 S_{\scriptscriptstyle E} \approx
 {2 c^2\lambda R_t \over  g^2 \cos 2\theta_{\scriptscriptstyle E}}
~ \bigg[1 - {\sin(2\lambda R_t)\over 2\lambda R_t}\bigg]~.
\label{eq:E4.6}\end{eqnarray} 
In order to understand the physical
mechanism that provides this Euclidean action, let us use the
representation of topological charge via the divergence of the
Chern-Simons current,
\begin{eqnarray}
Q ={1\over 4\pi^2} \oint d\sigma_\mu K^\mu = {1\over 4\pi^2} \int d^4x
~\partial_\mu K^\mu, \label{eq:E4.8}\end{eqnarray} where
\begin{eqnarray}
K^\mu ={1\over 4} \epsilon^{\mu\nu\rho\sigma} \big[A_\nu^a\partial_\rho
A_\sigma^a + {1\over 3}\epsilon_{abc}A_\nu^a A_\rho^b A_\sigma^c \big]~.
\label{eq:E4.9}\end{eqnarray}
The second term (usually the major one)
identically vanishes since the ephemeron field has only one color
component. It is easy to show, that the components of $K^\mu$ can be
presented in a compact form,
\begin{eqnarray}
K^\tau  = \pm {1\over 4} \sqrt{\rm g}~ {\rm g}^{lm} A_l E_m~,~~~ K^l =
-{1\over 4} \epsilon^{lmn}A_m E_n =
 {1\over 4} [{\bf E}\times {\bf A}]^l~,
\label{eq:E4.10}\end{eqnarray}
where ${\rm g}^{lm}={\rm
diag}[1,r^{-2},\tau^{-2}]$, $E_m=-\partial_\tau A_m$ is the electric
field strength, and $\epsilon^{lmn}$ is the three-dimensional Levi-Civita
tensor, $\epsilon^{r\phi\eta}=1$.\footnote{Deriving the expression for
$K^\tau$, we explicitly used the condition of self-duality; this is
where the alternating sign came from. The alternating sign in the spatial
components will come from the fields defined by Eq.(\ref{eq:E2.20}).} We
can see now, that the spatial components of the topological current
coincide with the canonical vector of the photon spin, thus being related
to the polarization properties of the ephemeron field. Using the
representation (\ref{eq:E4.9}), we can write down the topological charge
as a sum of the three surface integrals,
\begin{eqnarray}
Q ={1\over 2\pi} \bigg\{ \int_{-\pi/4}^{\pi/4} d\eta \int_{0}^{R_t} d r
\big[ K^\tau(\infty,r,\eta)- K^\tau(0,r,\eta)\big]\nonumber \\
+\int_{-\pi/4}^{\pi/4} d\eta \int_{0}^{\infty} d \tau \big[
K^r(\tau,R_t,\eta)- K^r(\tau,0,\eta)\big]\nonumber \\ + \int_{0}^{\infty}
d \tau  \int_{0}^{R_t} d r \big[ K^\eta(\tau,r,\pi/4)-
K^\eta(\tau,r,-\pi/4)\big] \bigg\}~. \label{eq:E4.11}\end{eqnarray} 
Next,
it is easy to show, that $K^\tau(\infty,r,\eta)=0$, $K^\tau(0,r,\eta)=0$,
and $K^r(\tau,0,\eta)=0$; this immediately follows from
Eq.~(\ref{eq:E4.10}) and the expressions for the field components.
Furthermore, for a large $\lambda R_t$, the contribution of $
K^r(\tau,R_t,\eta)$ is suppressed by an extra factor $(\lambda
R_t)^{-1}$. As a result, $Q$ is defined mainly by the last line in
Eq.~(\ref{eq:E4.11}). Thus, the topological charge and, by virtue of
(\ref{eq:E4.3}), the Euclidean action are defined by the difference of the
flux of the rapidity component $K^\eta$ of the canonical spin between the
hyper-planes $\eta_{\scriptscriptstyle E}=\pm\pi/4$
($\eta_{\scriptscriptstyle M}= \pm\infty$),
\begin{eqnarray}
Q \approx {c^2\lambda^2\over 8\pi} \int_{0}^{\infty} d \tau
\int_{0}^{R_t} d r \bigg[ \lambda r ~J_1^2(\lambda r)~\sin(\eta-\theta)
~e^{-\lambda\tau
\cos(\eta-\theta)}\bigg]_{\eta=-\pi/4}^{\eta=\pi/4}\nonumber\\ =
{c^2\lambda^2 R_t^2 \over 8\pi }~~
 {1 \over\cos 2\theta_{\scriptscriptstyle E}}~
 \bigg[J_0^2(\lambda R_t)+ J_1^2(\lambda R_t)
 - 2 {J_0(\lambda R_t) J_1(\lambda R_t)\over \lambda R_t}\bigg]~.
\label{eq:E4.12}\end{eqnarray}
This effect is clearly of a topological
nature. Indeed, by observation, the radial component (\ref{eq:E2.20})
$$A_r =R'_{\theta, \lambda,\vec{r}_0}(x_{\scriptscriptstyle E})\propto
\tan(\eta_{\scriptscriptstyle E}-\theta_{\scriptscriptstyle E})~,$$ of
the Euclidean field, or alternatively, the the radial component
(\ref{eq:E3.12}),
$$A_r=R'_{\theta,\lambda,\vec{r}_0}(x_{\scriptscriptstyle M})\propto
\tanh(\eta_{\scriptscriptstyle M}-\theta_{\scriptscriptstyle M})~,$$ of
the Minkowski field change their sign when the coordinate $\eta$ crosses
the rapidity center $\theta$ of the state. The function that describes
this transition is shaped exactly as a classical kink. Since the
component $A_\phi$ does not changes its sign, the observer that moves
along the $\eta$ direction will encounter a gradual change of the
circular polarization (direction of the photon spin) of the ephemeron
field.\footnote{One can easily recognize the similarity between the
polarization properties of the ephemeron field and those of the system of
two photons with opposite spins and moving in opposite directions. Such a
system emerges in the decay $\pi^0\to 2\gamma$. In that case, the
amplitude of the process includes an Abelian axial anomaly which is
proportional to the topological charge, though the field of the photons
is not self-dual, and the topological charge is not proportional to the
minimal action.} This is also seen directly from the spin density
$K^\eta$ ( the integrand in Eq.~(\ref{eq:E4.12})~). Thus, we encounter a
new example of Thomas precession, inherent in all states with spin in the
wedge dynamics. For the Dirac field, this effect has been discussed in
paper [II]. In both cases, the effect is due to the {\em curvature} of
the hypersurface of constant proper time $\tau$, along which the
states of the wedge dynamics are normalized by the system of the moving
observers. Indeed, in curvilinear coordinates, the spin connection $$
\omega_{\eta}^{\tau\eta}=
e_\alpha^{~\tau}e_\beta^{~\eta}\omega_{\eta}^{~\alpha\beta}=
{1\over\tau}\omega_{\eta}^{~03}~, $$ is proportional to the curvature
$\tau^{-1}$; it is large at small $\tau$, and vanishes at large $\tau$.
This means that  at large $\tau$, the local physics remains unaffected by
the curvature.

Finally, solely for pedagogical reasons, we shall try to connect the
existence of the non-vanishing (and not integer) topological charge $Q$
with some kind of tunneling. The first thing to do is to find a
``potential barrier''. In our case, the barrier emerges because in 
the wedge
dynamics any state is detected by a system of local observers moving with
different velocities. Therefore, a transfer along the surface of constant
$\tau$ is impossible without acceleration.\footnote{The reader should not
be confused by this ``motion in the space-like direction''. In Euclidean
space, all directions are space-like. Instantons also do not necessarily
provide tunneling between the configurations which are separated by a
time interval.} The physical effect of this acceleration is Thomas
precession, which rotates the spin of the state in opposite directions
when the state is detected on different sides of the rapidity center of
the wave packet. It is exactly the work done by the forces of inertia
which creates a potential barrier between these two degenerate states.
One can say, that this degeneracy is provided by the tunneling. 

\section{Conclusion}
\label{sec:S5}

The theory of ultrarelativistic nuclear collisions is a challenging field
of research where the price of correctly asked questions amounts to the
value of the practical results. { }From the perspective of the work
accomplished in the previous \cite{QFK}-\cite{fse} and the present
paper, we may claim that the scenario of ultrarelativistic nuclear
collisions, despite an enormous complexity of the phenomenon, can be at
least drafted from first principles. Its design must include two key
ingredients, dynamical definition of the ``final'' states as the
collective modes of the quark-gluon matter at the intermediate stage of
the evolution, and the mechanism responsible for the breakdown of the
nuclei coherence. The existence of the Euclidean ephemeron solutions
seems to resolve the issue of the nuclei decoherence exactly in the same way as
the instantons solve the problem of their integrity before the
collisions.  A Euclidean path through the instanton liquid provides the
least action to a confined group of quarks and make the color degrees of
freedom invisible. The coherence of the nuclear wave function is lost
because the path of the least Euclidean action includes new field
configurations which are the Euclidean images of the propagating states
and do not interlock the color and space directions. At the last instance
before the collision the nuclei just {\em become unbound}.

The physical meaning of ephemerons is exactly the same as the meaning
of instantons. Indeed, one can easily check that\\
~~(i)~ Since the Euclidean field strength tensor
$F^{\scriptscriptstyle (E)}_{\mu\nu}$ is self-dual, the Euclidean
energy-momentum tensor vanishes, $T^{\scriptscriptstyle (E)}_{\mu\nu}
(F^{\scriptscriptstyle (E)})\equiv 0$;\\
~~(ii)~ The analytic continuation $x_{\scriptscriptstyle E}\to
x_{\scriptscriptstyle M}$ leads to the $F^{\scriptscriptstyle
(M)}_{\mu\nu}$, in which $ B^k_{\scriptscriptstyle E}\to
B^k_{\scriptscriptstyle M}$, and $ E^k_{\scriptscriptstyle E}\to
i E^k_{\scriptscriptstyle M}$;\\
~~(iii)~ The energy-momentum tensor of the Minkowski metric computed
on these analytically continued fields vanishes also,
$T^{\scriptscriptstyle (M)}_{\mu\nu}
(F^{\scriptscriptstyle (M)})\equiv 0$.\\
This is a clear indication that the ephemerons indeed are the pure vacuum
fluctuations carrying no energy or momentum. In 
the Euclidean version of the
wedge dynamics, they must be included into the ensemble of the classical
field configurations on the same footing as instantons. A distinctive
feature of the wedge dynamics is the localization of the states with
a given velocity. Therefore, at large $\tau$, the wave packets
representing the two nuclei are well separated, and each of them can be
studied locally in its own rest frame. This property does not change in
the Euclidean wedge dynamics, which provides {\em a unique opportunity to
consider both the confining regime and the interaction of the two nuclei
in a common Euclidean projection}. If no interaction between the nuclei
occurs, then the ephemerons appear and vanish as the pure gauge fields
without any material trace. Only a real interaction,
that explicitly breaks the translational symmetry, can supply them with
the energy and excite them as the propagating fields.

The instanton physics, despite the global nature of mathematical theorems
related to the topological field configurations, is, as a matter of
fact, the {\em local physics} in isotropic  space. Therefore, we may
expect, that at $\tau\geq 1fm$ all physical mechanisms that stabilize the
instanton liquid \cite{ShuSch} are not corrupted, 
and the instantons of the
size $\rho_i < \tau$ do not have real competitors. The ephemerons with
small $\lambda$ (i.e. large size $\rho_e=\lambda^{-1}$), despite 
the fact that their
fields are weak, are not likely to survive, e.g., the long lasting
interactions with light quarks. Only the short-lived ephemerons of small
size are expected to be truly active, since  they show up only at the
latest moments before the nuclei intersect. Then, they have many
properties that one may wish to associate with  partons. One may also
see sufficient differences. They cannot be factorized out of the
fast-moving nuclei as the point-like color charges or the independent
plane waves. At the
earliest moments, they are widely extended in the rapidity direction and
localized in the transverse plane. They are ``made of'' the electric and
magnetic fields with highly non-trivial polarization. They are likely to
interact by chromo-magnetic rather  than by chromo-electric forces. They
are the fluctuations  that can be resolved in the course of 
a collision and
become free (with all reservations regarding the final-state
interactions) only after
the collision. It is still necessary to learn how the
ephemerons interact with quarks and between each other, and if there
exists a critical density of ephemerons in the transverse plane.

Regardless of the ephemeron's size, its Euclidean action rapidly grows 
when  $|\theta_{\scriptscriptstyle E}|\to\pi/4$,
$|\tan\theta_{\scriptscriptstyle E}|\to 1$. Then, 
$$ e^{-S_{\scriptscriptstyle E}}\propto \exp\bigg\{ -{\lambda R_t 
\over  g^2\cos 2\theta_{\scriptscriptstyle E}}\bigg\}~.$$ 
The weight $\exp[- S_{\scriptscriptstyle E}]$ strongly suppresses the
Euclidean fluctuations with the extreme rapidities
$\theta_{\scriptscriptstyle E}$. After analytic continuation, these
fluctuations would become the partons with the largest Minkowski
rapidities $\theta_{\scriptscriptstyle M}$ which, it their turn,
correspond to the smallest $x_{\scriptscriptstyle F} \sim
e^{-\theta_{\scriptscriptstyle M}}$.

If the one-instanton solution had not been discovered analytically more
than 20 years ago \cite{BPST}, it would 
have been found much later by means of
the lattice calculations as a constituent of multi-instanton
configurations. We suggest that the ephemeron solutions can also be found
from the lattice calculations if the periodic boundary conditions are
dropped and special attention is paid to the ``corner'' between
the $x^+xy$- and 
the $x^-xy$-hyperplanes. The problem must be posed in the gauge
$A^\tau=0$, and the boundary condition $A_\eta(\tau=0)=0$ must be imposed
on the gluon field. This type of lattice simulations would immediately
account for the interactions between ephemerons, ephemerons and
instantons, etc.

\bigskip

\noindent {\bf ACKNOWLEDGMENTS}

The author is grateful to Berndt Muller, Edward Shuryak and
Eugene Surdutovich for the helpful
conversations at various stages in the development of this work, and
appreciate the help of Scott Payson who critically read the manuscript.

\end{document}